\documentstyle[psfig,twocolumn,prb,aps]{revtex}

\begin{document}
\draft
\twocolumn[\hsize\textwidth\columnwidth\hsize\csname %
 @twocolumnfalse\endcsname

\title{Models of impurities in valence bond  spin chains and
ladders}

\author{Alexei K. Kolezhuk\protect\cite{perm}}
\address{Institut f\"{u}r Theoretische Physik,
Universit\"{a}t Hannover, Appelstr. 2, D-30167 Hannover, Germany\\
Institute of Magnetism, National Academy of 
Sciences and Ministry of Education of Ukraine\\
36(b) Vernadskii avenue, 252142 Kiev, Ukraine}

\date{March 27, 1998;  printed \today}

\maketitle

\begin{abstract}
We present the class of models of a nonmagnetic impurity in
$S={1\over2}$ generalized ladder with an AKLT-type valence bond ground
state, and of a $S={1\over2}$ impurity in the $S=1$ AKLT chain. The ground
state in presence of impurity can be found exactly. Recently studied
phenomenon of local enhancement of antiferromagnetic correlations
around the impurity is absent for this family of models.
\end{abstract}


]

Over the last decade, low-dimensional spin systems, particularly the
Heisenberg spin chains and ladders, have continued to attract
considerable attention of
researchers.\cite{DagottoRice96,affleck89rev} The interest to spin
ladders is particularly stimulated by the hope to get some insight
into the physics of metal-oxide superconductors; in support of this
hope, superconductivity in the ladder compound
Sr$_{0.4}$Ca$_{13.6}$Cu$_{24}$O$_{41.84}$ in presence of hole doping
and high pressure was recently reported.\cite{Uehara+96} It is now
well established that ``regular'' (i.e., with only ``leg'' and
``rung'' exchange couplings) $S={1\over2}$ isotropic spin ladders have
a disordered gapped ground state when the number of legs is even,
while odd-legged ladders have quasi-long-range ordered gapless ground
state.  On the other hand, ``generalized'' ladders including other
couplings can serve as interesting toy models with a rich behavior
which is often very different from that of ``regular'' models.
\cite{BG93+,White96,KM97-98,NT97,Weihong+,KM98xxx}

Recently, interesting experimental results on ladders doped with
nonmagnetic impurities (Cu substituted by Zn) have been
obtained:\cite{Azuma+97}  surprisingly, the antiferromagnetic (AF) order was
found to be stabilized by the doping; a similar behavior has also been
observed in spin-Peierls chains.\cite{CuGeO3} A number of numerical
studies\cite{Ng96+etc,Martins+97} indicated that local AF correlations near a
nonmagnetic impurity are enhanced comparing to the system without
vacancies. It has been suggested that this phenomenon, as well as
several other similar effects in one- and two-dimensional
antiferromagnets, \cite{Miyashita+etc} can be explained on a common basis
using the so-called ``pruned'' resonating valence bond (RVB)
picture.\cite{Martins+97} Nonmagnetic impurity affects
formation of instant singlet bonds for spins which are located in its
immediate vicinity, making some of the bonds geometrically impossible and
thus enhancing  the other bonds. This explanation is supposed to
be rather general and does not depend much on the interaction details.

In this paper I show that for certain models of nonmagnetic impurities
in generalized $S={1\over2}$ spin ladders with exact matrix-product
ground states of the type considered by us recently,
\cite{KM97-98,KM98xxx} local AF correlations are partly or 
completely insensitive to the presence of impurity. 

\begin{figure}
\mbox{\hspace{5mm}\psfig{figure=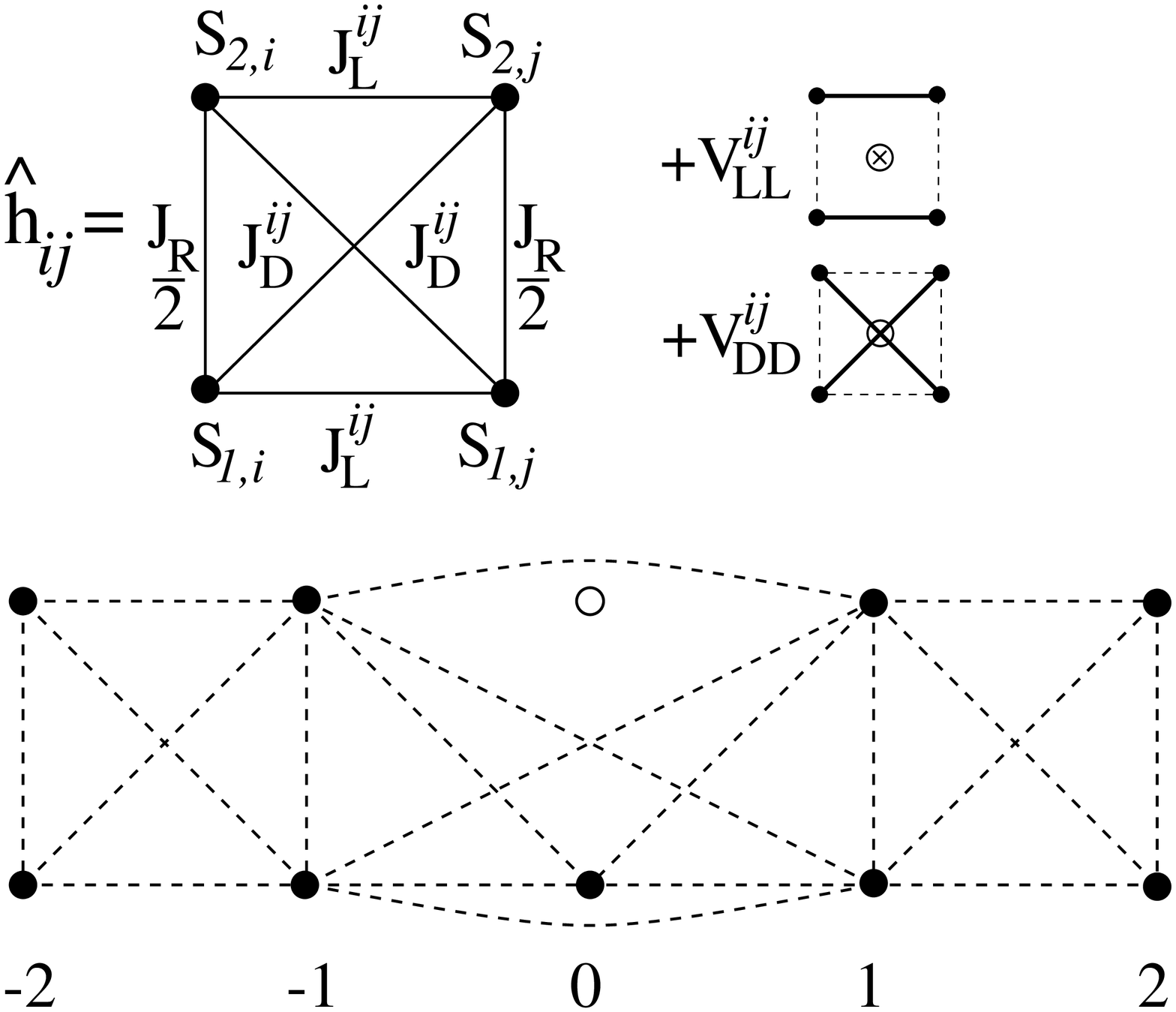,width=75mm,angle=-90}}
\vspace{3mm}
\caption{\label{fig:implad} Nonmagnetic impurity in the generalized
$S={1\over2}$ spin ladder as described by the Hamiltonian
(\protect\ref{ham}), $V$'s denote the biquadratic couplings.  }
\end{figure}

Consider the model of a vacancy in the generalized $S={1\over2}$
ladder with additional diagonal and biquadratic interactions,
described by the following Hamiltonian:
\begin{eqnarray} 
\label{ham} 
\widehat H&=&\sum_{i}
\widehat{h}_{i,i+1} +\widehat{h}_{-1,1}\,,\\
\widehat{h}_{ij}&=&
{1\over2}\,J_R ({\mathbf S}_{1i}{\cdot\mathbf S}_{2i}
  +  {\mathbf S}_{1,i+1}{\cdot\mathbf S}_{2,i+1})
\nonumber\\
&+& J_{L}^{ij}({\mathbf S}_{1i}{\cdot\mathbf S}_{1j} +
 {\mathbf S}_{2i}{\cdot\mathbf S}_{2j})
+J_{D}^{ij}({\mathbf S}_{1i} {\cdot\mathbf S}_{2j} 
+{\mathbf S}_{2i}{\cdot\mathbf S}_{1j})\nonumber\\
&+&V_{LL}^{ij} ({\mathbf S}_{1i}{\cdot\mathbf S}_{1j}) 
             ({\mathbf S}_{2i}{\cdot\mathbf S}_{2j})
+V_{DD}^{ij}({\mathbf S}_{1i}{\cdot\mathbf S}_{2j})
               ({\mathbf S}_{2i}{\cdot\mathbf S}_{1j})\,,\nonumber
\end{eqnarray}
here the indices $1$ and $2$ distinguish lower and upper legs, and $i$
labels rungs (see Fig.\ \ref{fig:implad}), and the terms involving the
vacancy site $S_{2,0}$ are implicitly assumed to be missing in
$\widehat{h}_{0,1}$ and $\widehat{h}_{-1,0}$. The ``bulk'' couplings
$J_{R}$, $J_{L}^{i,i+1}=J_{D}^{i,i+1}=1$, and
$V_{LL}^{i,i+1}=V_{DD}^{i,i+1}={4\over5}$ do not depend on $i$,
$J_{R}$ is a free parameter, and we have introduced the extra ``edge''
interaction between the rungs $-1$ and $1$
across the vacancy  to
make the problem solvable.

In absence of the vacancy the model (\ref{ham}) describes the
generalized Bose-Gayen model as introduced in Ref.\
\onlinecite{KM98xxx}, at the special value of the leg/diagonal
coupling ratio equal to $1$. At $J_{R}>{8\over5}$ its ground state is
a product of singlet bonds along the ladder rungs, and we will be
interested in the interval $J_{R}<{8\over5}$, where the ground state
coincides with that of the effective $S=1$ Affleck-Kennedy-Lieb-Tasaki
(AKLT) chain,\cite{AKLT} whose $S=1$ spins are formed by the triplet
degrees of freedom of the rungs. \cite{KM98xxx} This effective AKLT
ground state can be conveniently written in a form of the so-called
matrix product state: \cite{Fannes+,Klumper+}
\begin{equation} 
\label{aklt}
\Psi_{0} =\mbox{tr}(\prod_i  g_i),\quad
g_i  = {1\over\sqrt{3}} \left[ 
\begin{array}{lr}
| t_0\rangle_i & - \sqrt{2}  | t_{+} \rangle_i \\
\sqrt{2}  | t_{-} \rangle_i &  -  | t_0 \rangle_i
\end{array} 
\right],
\end{equation} 
where $|t_{\mu}\rangle_{i}$, $\mu=0,\pm1$ are the triplet states of the
$i$-th rung. The ground state energy per rung is \cite{KM98xxx}
\[
E_{0}=-13/10+J_{R}/4\,.
\]

\begin{figure}
\mbox{\hspace{5mm}\psfig{figure=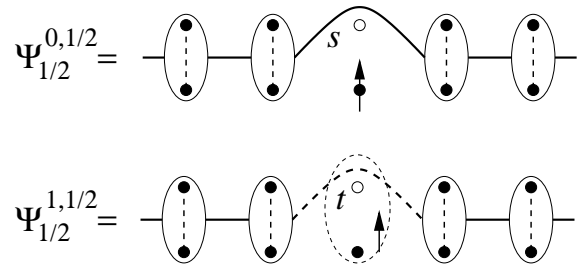,width=75mm,angle=-90}}
\vspace{3mm}
\caption{\label{fig:imp-vbs}
Schematic valence-bond representation of the wave functions
$\Psi^{0,1/2}_{1/2}$ and $\Psi^{1,1/2}_{1/2}$ contained in
(\protect\ref{ansatz}); solid and dashed lines denote singlet and
triplet valence bond links, respectively. Solid ovals indicate that
spins on each rung are coupled into effective triplet; dashed oval in
the bottom picture denotes that the triplet valence bond and the
unpaired spin are coupled into a spin-$1\over2$ state.
}
\end{figure}

We will look for the wave function of the ground state in
presence of the impurity in the form of the following matrix product:
\begin{equation} 
\label{ansatz} 
\Psi_{0}^{\text{imp}}=\mbox{tr}(g_{-N}\cdots g_{-1}G_{0}g_{1}\cdots g_{N})\,,
\end{equation}
where the matrix $G_{0}$ corresponding to the unpaired spin at the
$0$-th rung is chosen from the requirement that
$\Psi_{0}^{\text{imp}}$ has  both the total spin and its $z$-projection
equal to ${1\over2}$; the most general form of $G_{0}$ is\cite{KMY97}
\begin{equation} 
\label{Gimp} 
G_{0}={1\over\sqrt{3+x^{2}}}\left[ 
\begin{array}{lr}
(x-1)| \uparrow\rangle & 0 \\
-2 |  \downarrow\rangle &  (x+1) | \uparrow \rangle
\end{array} 
\right],
\end{equation}
$x$ being a free parameter. Physically, the wave function
$\Psi_{0}^{\text{imp}}$ describes a superposition
$(x/\sqrt{3})\Psi^{0,1/2}_{1/2}+\Psi^{1,1/2}_{1/2}$, where
$\Psi^{j_{\text{lad}},1/2 }_{j_{\text{tot}}}$ denotes a
wave function with the total spin $j_{\text{tot}}$ composed from the
states of the unpaired spin ${1\over2}$ and the states of the rest of the
ladder having total spin $j_{\text{lad}}$; in $\Psi^{0,1/2}_{1/2}$ the
unpaired spin is completely decoupled from the rest of the ladder
(which is in the effective AKLT state with one valence bond across the
impurity),
while in $\Psi^{1,1/2}_{1/2}$ it is coupled with the edge
Kennedy-Tasaki triplet \cite{KennedyTasaki92} into  the state with
$j_{\text{tot}}={1\over2}$, see Fig.\ \ref{fig:imp-vbs}.

Further, the Hamiltonian (\ref{ham}) conserves parity with respect to
the mirror transformation $i\mapsto -i$, and one can see that
$\Psi^{1,1/2}_{1/2}$ and $\Psi^{0,1/2}_{1/2}$ have different
parities. The solution with completely decoupled unpaired spin is not
very interesting, so that we will look for the ground state wave
function of the form (\ref{ansatz}), (\ref{Gimp}) with $x=0$. 
Following the approach outlined in Ref.\  \onlinecite{Klumper+}, we 
 demand that the local Hamiltonian
$\widehat{h}_{\text{imp}}$, defined as (recall that the terms with
$S_{2,0}$ should be dropped)
\begin{equation} 
\label{hint} 
\widehat{h}_{\text{imp}}=\widehat{h}_{-1,0}+\widehat{h}_{0,1} 
+\widehat{h}_{-1,1} -\varepsilon_{0}\,,
\end{equation}
where $\varepsilon_{0}$ is a free parameter, annihilates all states
contained in the matrix product $g_{-1}G_{0}g_{+1}$, and that all
other eigenstates of $\widehat{h}_{\text{imp}}$ have positive
energies. These conditions are sufficient for $\Psi_{0}^{\text{imp}}$
to be the ground state of $\widehat{H}$. The construction routine is
well described in literature, \cite{Klumper+,KM97-98,KM98xxx} so that I
only briefly address it here. 

The states of the $[-1,0,1]$ block can
be classified into multiplets $\Psi_{jm}$, where $j$ is the total spin
of the block and $m$ is its $z$-projection.
 In total, there are ten multiplets (five with
$j={1\over2}$, four with $j={3\over2}$, and one with $j={5\over2}$);
one can however straightforwardly check that the matrix product
$g_{-1}G_{0}g_{+1}$ contains only states of the following three 
multiplets:
\begin{equation} 
\label{lgs}
\Psi^{g,1}_{{1\over2},m}=\psi^{111}_{{1\over2},m}, \quad 
\Psi^{g,2}_{{1\over2},m}=\psi^{110}_{{1\over2},m},\quad 
\Psi^{g}_{{3\over2},m}=\psi^{112}_{{3\over2},m}\,,
\end{equation}
here $\psi^{S_{A},S_{B},S_{AB}}_{jm}$ denotes the state of the
$[-1,0,1]$ block with the total spin $j$, $S_{A}$, $S_{B}$, and
$S_{AB}$ being the total momenta of the $-1$th, $+1$th rung and the
$[-1,1]$ block, respectively. The local Hamiltonian
$\widehat{h}_{\text{imp}}$ should annihilate the states (\ref{lgs}),
so that it can be generally written as a projector onto the subspace
of the remaining seven multiplets,
\begin{eqnarray} 
\label{les} 
&&\Psi^{e,1}_{{1\over2},m}={1\over\sqrt{2}} 
(\psi^{101}_{{1\over2},m}+\psi^{011}_{{1\over2},m}),\quad
\Psi^{e,2}_{{1\over2},m}=\psi^{000}_{{1\over2},m},\nonumber\\
&& \Psi^{e,3}_{{1\over2},m}={1\over\sqrt{2}} 
(\psi^{101}_{{1\over2},m}-\psi^{011}_{{1\over2},m}),\\
&&\Psi^{e,1}_{{3\over2},m}={1\over\sqrt{2}} 
(\psi^{101}_{{3\over2},m}+\psi^{011}_{{3\over2},m}),\quad
\Psi^{e,2}_{{3\over2},m}=\psi^{111}_{{3\over2},m},\nonumber\\
&& \Psi^{e,3}_{{3\over2},m}={1\over\sqrt{2}} 
(\psi^{101}_{{3\over2},m}-\psi^{011}_{{3\over2},m}),\quad 
\Psi^{e}_{{5\over2},m}=\psi^{112}_{{5\over2},m}\,.\nonumber
\end{eqnarray}
We make a further simplification, assuming that
$\widehat{h}_{\text{imp}}$ does not mix the above multiplets, so that
\begin{eqnarray} 
\label{ham-proj} 
\widehat{h}_{\text{imp}}=
\sum_{j={1\over2},{3\over2},{5\over2}}\sum_{i} 
\sum_{m=-j}^{j}\lambda_{j}^{(i)}
|\Psi^{e,i}_{jm}\rangle
\langle \Psi^{e,i}_{jm}| \,,
\end{eqnarray}
where all $\lambda_{j}^{(i)}$ should be positive to ensure that
(\ref{ansatz}) is the ground state.  Demanding further that this
structure is compatible with the particular form of the Hamiltonian
(\ref{ham}), one arrives at the following family of solutions for the
coupling constants and the parameter $\varepsilon_{0}$:
\begin{eqnarray} 
\label{sol} 
&& J_{L}^{-1,1}=\lambda_{1/2}^{(1)}/2 +(1+J_{R})/4, \;\;
 V_{LL}^{-1,1}=J_{R}+2\lambda_{1/2}^{(1)}-1,\nonumber\\
&&
J_{D}^{-1,1}=(1-\lambda_{1/2}^{(1)})/2-J_{R}/4,
\quad
V_{DD}^{-1,1}=-2\lambda_{1/2}^{(1)}-J_{R},\nonumber\\
&& \varepsilon_{0}=-19/16+ J_{R}/4\,,
\end{eqnarray}
where $\lambda_{1/2}^{(1)}$ plays the role of a free  parameter, and the
expressions for the other eigenvalues are
\begin{eqnarray} 
\label{lambdas} 
&& \lambda_{1/2}^{(2)}=1-J_{R},\quad
\lambda_{1/2}^{(3)}=1/4-(\lambda_{1/2}^{(1)}+J_{R})/2,\nonumber\\ 
&& \lambda_{3/2}^{(1)}=3/2+\lambda_{1/2}^{(1)},\quad
\lambda_{3/2}^{(2)}=3/2,\\
&&\lambda_{3/2}^{(3)}=2/5-(\lambda_{1/2}^{(1)}+J_{R})/5,
\quad \lambda_{5/2}=5/2.\nonumber
\end{eqnarray}
The parameter $\varepsilon_{0}$ has the meaning of a ground state
energy of the $[-1,0,1]$ block, thus the states with and without a
vacancy differ in energy by the value $\varepsilon_{0}-2E_{0}$.  The
conditions of positivity of $\widehat{h}_{\text{imp}}$ require that
\begin{equation} 
\label{cond} 
J_{R}\leq {1\over2}-\lambda_{1/2}^{(1)},\quad \lambda_{1/2}^{(1)}\geq 0\,.
\end{equation}
The most symmetric solution from the above family is achieved by
setting $\lambda_{1/2}^{(1)}=(1-2J_{R})/4$, $J_{R}\leq 1/2$, then
$J_{L}^{-1,1}= J_{D}^{-1,1}=3/8$ and
$V_{LL}^{-1,1}=V_{DD}^{-1,1}=-1/2$. 

Using the standard matrix product technique, it is easy to calculate
the spin correlation functions and distribution of the excess spin  in the
state (\ref{ansatz}). The mean value of $S^{z}$ at each site is given
by
\begin{equation} 
\label{excess} 
\langle S_{1,0}^{z}\rangle =-{1\over6},\quad
\langle S_{1,i}^{z}\rangle=\langle
S_{2,i}^{z}\rangle={2\over9}\left(-{1\over3}\right)^{|i|-1}\,,
\end{equation}
here $|i|\geq 1$. Following Ref.\ \onlinecite{Martins+97}, we
calculate the spin correlation functions along the ladder legs, with
the starting site being next to the vacancy, and compare them to the
correlations in absence of the vacancy. Quite surprisingly, one finds
that the AF correlations are not at all affected by the presence of a
vacancy:
\begin{eqnarray} 
\label{afcorr}
 \langle S_{1,0}^{\alpha} S_{1,i}^{\alpha}\rangle &=& \langle
S_{2,1}^{\alpha} S_{2,i+1}^{\alpha}\rangle
=(-1)^{|i|}\cdot3^{-|i|-1} \nonumber\\ 
&=& \langle S_{1,n}^{\alpha}S_{1,n+i}^{\alpha}\rangle_{\text{w/o}}
=\langle S_{2,n}^{\alpha}S_{2,n+i}^{\alpha}\rangle_{\text{w/o}}\,;
\end{eqnarray}
here $\alpha=x,y,z$, ``w/o'' in the second line means ``without
vacancy,'' and $|i|\geq 1$. Note that despite the presence of the
excess spin,  the spin correlations remain {\em isotropic.\/}

One can show that the above features (insensitivity of AF
correlations to the presence of a vacancy and their isotropic
character) survive also in more complicated matrix-product-solvable
models: the ansatz (\ref{ansatz}), (\ref{Gimp}) can be obviously used
in its most 
general form, with the ``bulk'' matrices $g_{i}$ including singlet
degrees of freedom of the ladder rungs,\cite{BMN96}
\[
g_{i}\mapsto g_i(u)  = {1\over\sqrt{3+u^{2}}} \left[ 
\begin{array}{lr}
u | s \rangle_i + | t_0\rangle_i & - \sqrt{2}  | t_{+} \rangle_i \\
\sqrt{2}  | t_{-} \rangle_i & u | s \rangle_i -  | t_0 \rangle_i
\end{array} 
\right],
\]
then generally the Hamiltonian (\ref{ham}) will not be invariant under
the parity transformation $i\to -i$, so that the matrix (\ref{Gimp})
can be also used in its general form with arbitrary $x$.  [The
corresponding family of impurity models with exact ground states can
be obtained for this general ansatz, exactly in the same way as we did
above; however, the resulting model Hamiltonians are extremely
cumbersome and therefore we do not present those solutions here.]  For
such a state, the above formula for the correlations will change as
follows:
\begin{eqnarray} 
 \label{afcorr-gen}
\langle
S_{1,1}^{\alpha} S_{1,i+1}^{\alpha}\rangle&=&\langle
S_{2,1}^{\alpha} S_{2,i+1}^{\alpha}\rangle=
\langle S_{1,n}^{\alpha}S_{1,n+i}^{\alpha}\rangle_{\text{w/o}}\nonumber\\
&=&\langle S_{2,n}^{\alpha}S_{2,n+i}^{\alpha}\rangle_{\text{w/o}}=
q^{|i|}/(u^{2}+3)\,,\\
\langle S_{1,0}^{\alpha} S_{1,i}^{\alpha}\rangle &=&
{(1+x) q^{|i|}\over (1-u)(3+x^{2})},\quad q={u^{2}-1\over
u^{2}+3}\,.\nonumber 
\end{eqnarray}
One can see that the AF correlations along the legs are not affected,
except for the correlations involving the unpaired spin $S_{1,0}$,
which are enhanced for $u$ being in the interval between $-x$ and
$(x-3)/(1+x)$ and suppressed otherwise.  In valence-bond-type models
the decay of all correlations is purely exponential for all distances,
and presence of the impurity can only change the prefactor in front of
the exponent; accidentally, for the chosen ansatz
(\ref{ansatz},\ref{Gimp}) the changes coming from the excess spin and
``distortions'' due to the presence of a vacancy completely compensate
each other.  The spin excess distribution is also modified and is
generally asymmetric:
\begin{eqnarray*} 
\langle S_{1,i}^{z}\rangle ={2(\sigma x-1) q^{|i|}
\over(1-\sigma u)(3+x^{2})},\quad && 
\langle S_{2,i}^{z}\rangle
={2(x-\sigma)q^{|i|} \over(u+\sigma)(3+x^{2})},\nonumber\\
\langle S_{1,0}^{z}\rangle={(x^{2}-1)\over 2(3+x^{2})},\quad &&
\sigma\equiv\mbox{sgn}(i),\quad |i|\geq 1\,. 
\end{eqnarray*}

\begin{figure}
\mbox{\hspace{5mm}\psfig{figure=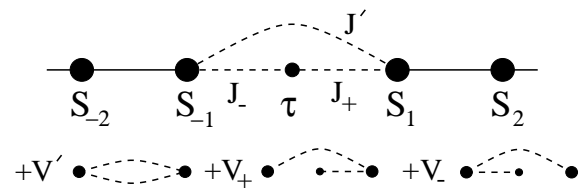,width=75mm,angle=-90}}
\vspace{3mm}
\caption{\label{fig:imp-aklt} $S={1\over2}$ impurity in the $S=1$ AKLT
chain as described by the Hamiltonian (\protect\ref{imp-aklt-ham}),
\protect\boldmath$\tau$\protect\unboldmath\ is the impurity spin, and
$V$'s indicate the biquadratic couplings.  
}
\end{figure}

Finally, one can observe that the model of a vacancy in the 
$S={1\over2}$ ladder can be reformulated as a
model of the $S={1\over2}$ impurity in the $S=1$ AKLT chain. Consider
the model described
by the following Hamiltonian:
\begin{eqnarray} 
\label{imp-aklt-ham}
&& \widehat{H}= \sum_{i\geq 1} (\widehat{h}^{\text{AKLT}}_{i,i+1}
 +\widehat{h}^{\text{AKLT}}_{-i,-i-1}) + \widehat{h}_{\text{imp}},\\
&& \widehat{h}_{\text{imp}}= (J_{+}{\mathbf S}_{1}
+J_{-}{\mathbf S}_{-1})\cdot
 \mbox{\boldmath$\tau$\unboldmath}
+J'({\mathbf S}_{-1}\cdot {\mathbf S}_{1})-\varepsilon_{0} \nonumber\\
&&\quad +({\mathbf S}_{-1}\cdot
 {\mathbf S}_{1})\big\{ 
(V_{+} {\mathbf S}_{1}
+V_{-}{\mathbf S}_{-1})\cdot
 \mbox{\boldmath$\tau$\unboldmath}\big\}
+ V' ({\mathbf S}_{-1}\cdot
 {\mathbf S}_{1})^{2}\,,\nonumber
\end{eqnarray}
here $\widehat{h}^{\text{AKLT}}_{i,j}={\mathbf S}_{i}\cdot {\mathbf
S}_{j} + {1\over3} ({\mathbf S}_{i}\cdot {\mathbf S}_{j})^{2}$ is the
local Hamiltonian of the AKLT chain in the bulk, and
$\widehat{h}_{\text{imp}}$ describes the interaction induced by 
presence of the impurity spin $\mbox{\boldmath$\tau$\unboldmath}$, see
Fig.\ \ref{fig:imp-aklt}, the parameter $\varepsilon_{0}$ being just a
constant energy shift having the meaning of the ground state energy
of the $[-1,\tau,1]$ block. Using the ansatz (\ref{ansatz}),
(\ref{Gimp}), one can repeat the entire construction routine as
described above for the ladder, and obtain the following family of
Hamiltonians for which $\Psi_{0}^{\text{imp}}$ is the exact ground state:
\begin{eqnarray} 
\label{sol-aklt} 
J_{\pm}&=&\{5(3-x^{2})/9 \pm 2x\}\lambda_{3/2}+5\lambda_{5/2}/9,\nonumber\\
J'&=& -(5+x^{2})\lambda_{3/2}/3 +5\lambda_{5/2}/6, \\
V_{\pm}&=&\{ -5(3+x^{2})/9 \pm 4x/3\}\lambda_{3/2}+5\lambda_{5/2}/9,\nonumber\\
V'&=&-(15+x^{2})\lambda_{3/2}/9 + 5\lambda_{5/2}/18,\nonumber\\
\varepsilon_{0}&=&(30-2x^{2})\lambda_{3/2}/9+5\lambda_{5/2}/9.\nonumber
\end{eqnarray}
Here $\lambda_{3/2}\geq 0$, $\lambda_{5/2}\geq 0$ are the eigenvalues
of $\widehat{h}_{\text{imp}}$ corresponding to the multiplets
\[
\Psi^{e}_{{3\over2},m}=(5+x^{2})^{-1/2} 
(x\psi^{112}_{{3\over2},m}-\sqrt{5}\psi^{111}_{{3\over2},m}),\quad
\Psi^{e}_{{5\over2},m}=\psi^{112}_{{5\over2},m},
\]
respectively. For $x\not=0,\infty$ the Eqs.\ (\ref{sol-aklt}) describe
models with an asymmetric impurity.  Again, as in case of the ladder,
one can straightforwardly check that the ``edge'' spin correlation
function in presence of the impurity $\langle S_{1}^{\alpha}
S_{i}^{\alpha}\rangle$ just coincides with that in the bulk,
independently of the value of $x$.  The above family contains two
interesting solutions: one is achieved by setting $x=0$,
$\lambda_{5/2}=3\lambda_{3/2}$ and describes the simple symmetric
model without biquadratic terms involving the impurity spin \boldmath$\tau$\unboldmath:
\begin{equation} 
\label{aklt1} 
J_{\pm}=J, \quad J'=-V'={1\over4} J,\quad V_{\pm}=0\, .
\end{equation}
Another solution corresponds to $\lambda_{3/2}=0$, then the ground
state of the model is twofold degenerate since both even and
odd-parity wave functions $\Psi_{0}^{\text{imp}}(x=0)$ and
$\Psi_{0}^{\text{imp}}(x=\infty)$ are eigenstates with the same
energy.

\smallskip

{\em Acknowledgments.---\/} I would like to thank E. Dagotto for
inspiring suggestions and critical comments, and H.-J. Mikeska for
discussion of the results. The hospitality of Hannover Institute for
Theoretical Physics is gratefully acknowledged.  This work was
supported by the German Ministry for Research and Technology (BMBF)
under the contract 03MI4HAN8 and by the Ukrainian Ministry of Science
(grant 2.4/27).

\end{document}